\documentclass[aps,prl,twocolumn,superscriptaddress,altaffilletter,nobibnotes,nofootinbib,10pt,showpacs,showkeys]{revtex4-1}

\def\pt{\mbox{$p_{\rm t} $}}   
   
\def\snn{\mbox{$\sqrt{s_{_{\rm NN}}}$}}   
 
\def \pb {Pb--Pb}
\def \au {Au--Au}

\newcommand{ \be }{\begin{eqnarray}}
\newcommand{ \ee }{\end{eqnarray}}
\newcommand{ \la }{\langle}
\newcommand{ \ra }{\rangle}

\newcommand{ \mean }[1]{\la #1 \ra}

\newcommand{ \psirp }{\Psi_{R}}

\usepackage{graphicx}  
\usepackage{dcolumn}   
\usepackage{bm}        
\usepackage{amssymb}   
\usepackage{lineno}
\usepackage{draftcopy}

\usepackage{color}
\definecolor{dgreen}{cmyk}{1.,0.,1.,0.2}        
\definecolor{orange}{cmyk}{0.,0.353,1.,0.}    

\usepackage[pdftex,bookmarks]{hyperref}

\usepackage{preprintcover}

\PreprintIdNumber{CERN-PH-EP-2010-059}
\PreprintCoverPaperTitle{Elliptic flow of charged particles in Pb--Pb collisions  
at $\sqrt{s_{\rm NN}} = 2.76$ TeV}
\PreprintCoverAbstract{We report the first measurement of charged particle elliptic flow in
Pb--Pb collisions at $\snn~=~2.76$~TeV with the ALICE detector at the
CERN Large Hadron Collider.  The measurement is performed in the
central pseudorapidity region $(|\eta|<0.8)$ and transverse momentum
range $0.2<p_{\rm t}<5.0$~GeV/$c$.  The elliptic flow signal $v_2$,
measured using the 4-particle correlation method, averaged over
transverse momentum and pseudorapidity is 0.087 $\pm$ 0.002 (stat)
$\pm$ 0.003 (syst) in the 40--50\% centrality class.  The differential
elliptic flow $v_2(p_{\rm t})$ reaches a maximum of 0.2 near $p_{\rm
  t} = 3$~GeV/$c$.  Compared to RHIC Au--Au collisions at $\snn~=
200$~GeV, the elliptic flow increases by about 30\%.  
Some
hydrodynamic model predictions which include viscous corrections are
in agreement with the observed increase.
}
\PreprintJournalName{PRL}

\begin{document}

\title{Elliptic flow of charged particles in Pb--Pb collisions  
at $\sqrt{s_{\rm NN}} = 2.76$~TeV}

%
\collaboration{The ALICE Collaboration} 
\noaffiliation

\author{K.~Aamodt}
\altaffiliation{}
\affiliation{Department of Physics and Technology, University of Bergen, Bergen, Norway}
\author{B.~Abelev}
\altaffiliation{}
\affiliation{Lawrence Livermore National Laboratory, Livermore, California, United States}
\author{A.~Abrahantes~Quintana}
\altaffiliation{}
\affiliation{Centro de Aplicaciones Tecnol\'{o}gicas y Desarrollo Nuclear (CEADEN), Havana, Cuba}
\author{D.~Adamov\'{a}}
\altaffiliation{}
\affiliation{Nuclear Physics Institute, Academy of Sciences of the Czech Republic, \v{R}e\v{z} u Prahy, Czech Republic}
\author{A.M.~Adare}
\altaffiliation{}
\affiliation{Yale University, New Haven, Connecticut, United States}
\author{M.M.~Aggarwal}
\altaffiliation{}
\affiliation{Physics Department, Panjab University, Chandigarh, India}
\author{G.~Aglieri~Rinella}
\altaffiliation{}
\affiliation{European Organization for Nuclear Research (CERN), Geneva, Switzerland}
\author{A.G.~Agocs}
\altaffiliation{}
\affiliation{KFKI Research Institute for Particle and Nuclear Physics, Hungarian Academy of Sciences, Budapest, Hungary}
\author{S.~Aguilar~Salazar}
\altaffiliation{}
\affiliation{Instituto de F\'{\i}sica, Universidad Nacional Aut\'{o}noma de M\'{e}xico, Mexico City, Mexico}
\author{Z.~Ahammed}
\altaffiliation{}
\affiliation{Variable Energy Cyclotron Centre, Kolkata, India}
% ...

\begin{abstract}
We report the first measurement of charged particle elliptic flow in
Pb--Pb collisions at $\snn~=~2.76$~TeV with the ALICE detector at the
CERN Large Hadron Collider.  The measurement is performed in the
central pseudorapidity region $(|\eta|<0.8)$ and transverse momentum
range $0.2<p_{\rm t}<5.0$~GeV/$c$.  The elliptic flow signal $v_2$,
measured using the 4-particle correlation method, averaged over
transverse momentum and pseudorapidity is 0.087 $\pm$ 0.002 (stat)
$\pm$ 0.003 (syst) in the 40--50\% centrality class.  The differential
elliptic flow $v_2(p_{\rm t})$ reaches a maximum of 0.2 near $p_{\rm
  t} = 3$~GeV/$c$.  Compared to RHIC Au--Au collisions at $\snn~=
200$~GeV, the elliptic flow increases by about 30\%.  
Some
hydrodynamic model predictions which include viscous corrections are
in agreement with the observed increase.
\end{abstract}

\maketitle

The goal of ultra-relativistic nuclear collisions is the creation and
study of the quark-gluon plasma (QGP), a state of matter whose
existence at high energy density is predicted by Quantum
Chromodynamics.  One of the experimental observables that is sensitive
to the properties of this matter is the azimuthal distribution of
particles in the plane perpendicular to the beam direction. When
nuclei collide at finite impact parameter (non-central collisions),
the geometrical overlap region and therefore the initial matter
distribution is anisotropic (almond shaped).  If the matter is
interacting, this spatial asymmetry is converted via multiple
collisions into an anisotropic momentum
distribution~\cite{Ollitrault:1992bk}.  The second moment of the
final state hadron azimuthal distribution is called elliptic flow; it
is a response of the dense system to the initial conditions and
therefore sensitive to the early and hot, strongly interacting phase
of the evolution. 

At RHIC large elliptic flow has been observed and is one of the key
experimental
discoveries~\cite{Ackermann:2000tr,Arsene:2004fa,Back:2004je,Adams:2005dq,Adcox:2004mh}.
Theoretical models, based on ideal relativistic hydrodynamics with a
QGP equation of state and zero shear viscosity, fail to describe
elliptic flow measurements at lower energies but describe RHIC data
reasonably well~\cite{Kestin:2008bh}.  Theoretical arguments, based on
the AdS/CFT conjecture~\cite{Maldacena:1997re}, suggest a universal
lower bound of $1/4\pi$~\cite{Kovtun:2004de} for the ratio of shear
viscosity to entropy density.  Recent model studies incorporating
viscous corrections indicate that the shear viscosity at RHIC is
within a factor of $\sim$5 of this
bound~\cite{Teaney:2003kp,Romatschke:2007mq,Masui:2009pw,Song:2008hj}.

The pure hydrodynamic models~\cite{Niemi:2008ta,Kestin:2008bh,Luzum:2009sb} 
and models which combine hydrodynamics with a hadron cascade afterburner 
(hybrid models)~\cite{Hirano:2010jg,Hirano:2005xf} that successfully describe
flow at RHIC predict an increase of the elliptic flow at the LHC
ranging from 10\% to 30\%, with the largest increase predicted by
models which account for viscous
corrections~\cite{Luzum:2009sb,Hirano:2010jg,Hirano:2005xf,Drescher:2007uh}
at RHIC energies.  
In models with viscous corrections,
  $v_2$ at RHIC is below the ideal hydrodynamic
  limit~\cite{Hirano:2005xf,Masui:2009pw} and therefore can 
show a stronger increase with energy.  
In hydrodynamic models the charged
particle elliptic flow as a function of transverse momentum does not
change significantly~\cite{Niemi:2008ta,Kestin:2008bh}, while the
$p_{\rm t}$-integrated elliptic flow increases due to the rise in
average $p_{\rm t}$ expected from larger radial (azimuthally
symmetric) flow.  The larger radial flow also leads to a decrease of
the elliptic flow at low transverse momentum, which is most pronounced
for heavy particles.  Models based on a parton
cascade~\cite{Molnar:2007an}, including models that take into account
quark recombination for particle production~\cite{Krieg:2007sx},
predict a stronger decrease of the elliptic flow as function of
transverse momentum compared to RHIC energies.  Phenomenological
extrapolations~\cite{Busza:2007ke} and models based on final state
interactions~\cite{Capella:2007kf} that have been tuned to describe
the RHIC data, predict an increase of the elliptic flow of $\sim50$\%,
larger than other models.  A measurement of elliptic flow at the LHC
is therefore crucial to test the validity of a hydrodynamic
description of the medium and to measure its thermodynamic properties,
in particular shear viscosity and the equation of
state~\cite{Abreu:2007kv}.

The azimuthal dependence of the particle yield can be written in the
form of a Fourier series~\cite{Voloshin:1994mz,Poskanzer:1998yz}:
\begin{equation}
E\frac{d^3 N}{d^3 p} = \frac{1}{2\pi} \frac{d^2 N}{p_{\rm t} dp_{\rm t}
  dy}\! \left(1\!+\!\sum_{n=1}^{\infty}2v_n \cos[n\!\left(
\phi\!-\!\Psi_R\right) ] \!\right), 
\label{eqFourier}
\end{equation}
where $E$ is the energy of the particle, $p$ the momentum, $p_{\rm t}$
the transverse momentum, $\phi$ the azimuthal angle, $y$ the rapidity,
and $\Psi_R$ the reaction plane angle.  The reaction plane is the
plane defined by the beam axis $z$ and the impact parameter
direction.  In general the coefficients $v_n = \mean{\cos[ n
    (\phi-\psirp)]}$ are $p_{\rm t}$ and $y$ dependent -- therefore we
refer to them as {\it differential flow}.  The {\it integrated flow}
is defined as an average evaluated with $d^2N/dp_{\rm t} dy$ used as a
weight.  The first coefficient, $v_1$, is called {\it directed flow},
and second coefficient, $v_2$, is called {\it elliptic flow}.  

We report the first measurement of elliptic flow of charged
particles in Pb--Pb collisions at the center of mass energy per nucleon pair 
$\snn~=~2.76$~TeV, with the ALICE
detector~\cite{Aamodt:2008zz,Carminati:2004fp,Alessandro:2006yt}.
The data were recorded in November 2010 during
the first run with heavy ions at the LHC.

For this analysis the ALICE Inner Tracking System (ITS) and the Time
Projection Chamber (TPC) were used to reconstruct the charged particle tracks. 
The VZERO counters and the Silicon Pixel Detector (SPD) 
were used for the trigger. 
The VZERO counters are two scintillator arrays providing both amplitude and 
timing information, covering the pseudorapidity range $2.8 < \eta < 5.1$ 
(VZERO-A) and $-3.7 < \eta < -1.7$ (VZERO-C).  
The VZERO time resolution is better than 1 ns.
The SPD is the innermost part of the ITS, consisting of two cylindrical layers of hybrid
silicon pixel assemblies  covering the range of $|\eta|<2.0$ and $|\eta|<1.4$ for the inner
and outer layer, respectively.
The minimum-bias interaction trigger required at least two out of the following three
conditions~\cite{alice_mult}: i) two pixel chips hit in the
outer layer of the silicon pixel detectors ii) a signal in VZERO-A iii) a signal in VZERO-C.
The bunch intensity was typically
$10^7$ Pb ions per bunch and each beam had 4 colliding bunches.  
The estimated luminosity was 
$5 \times 10^{23}$~cm$^{-2}$~s$^{-1}$, producing 
collisions with a minimum bias trigger at a rate of 50 Hz including about 
4 Hz nuclear interactions, 45 Hz
electromagnetic processes and 1 Hz beam background.  

\begin{figure}[thb]
 \begin{center}
   \includegraphics[width=0.49\textwidth]   {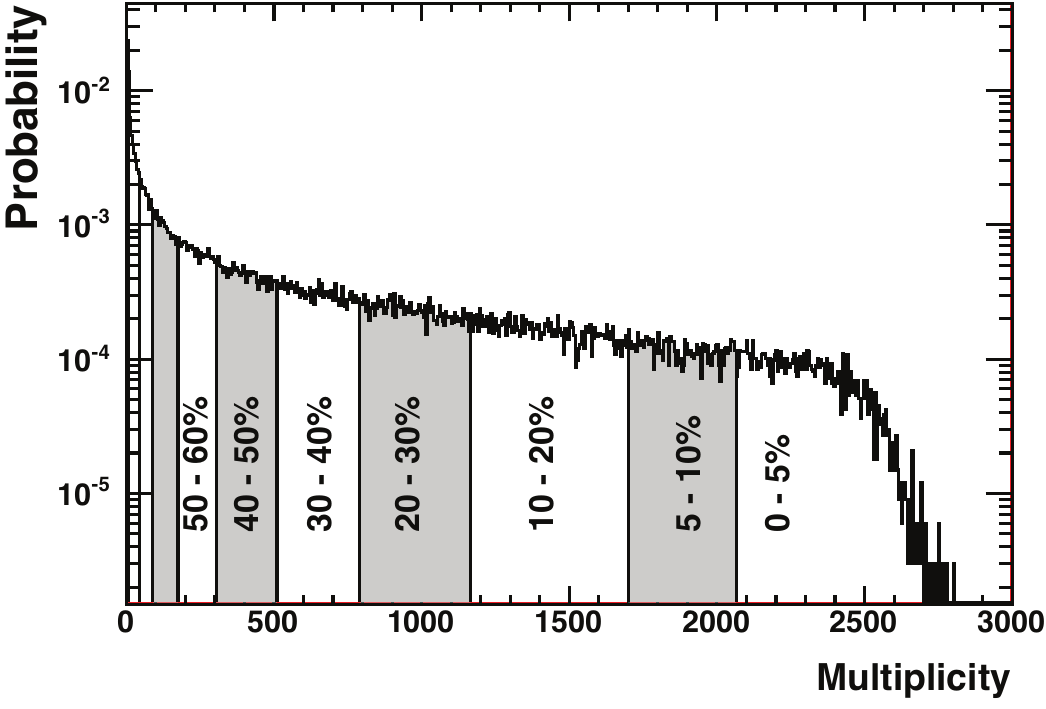}
   \caption{ The uncorrected multiplicity distribution of charged
     particles in the TPC $(|\eta|<0.8)$.  The
     centrality bins used in the analysis are shown and the cumulative
     fraction of the total events is indicated in percent. 
     The bins 60--70\% and 70--80\% are not labeled.}
    \label{centrality}
 \end{center}
\end{figure}

A removal of background events was carried out offline 
using the VZERO timing information and  
the requirement of two tracks in the central detector.
A study based on Glauber model fits to the multiplicity 
distribution (see also~\cite{alice_mult}) in the region corresponding to 80\% 
of the most central collisions, where the vertex reconstruction
is fully efficient, allows for the determination of the cross section 
percentile. Only events
with a vertex found in $|z| < 10$~cm were used in this analysis
to ensure a uniform acceptance in the central pseudorapidity region
$|\eta|<0.8$.  
An event sample of $45 \times 10^3$ Pb--Pb collisions passed the selection criteria and 
was used in this analysis. 
The data are analyzed in centrality classes determined by cuts on the 
uncorrected charged particle multiplicity, 
in pseudorapidity acceptance $|\eta|<0.8$.
Figure~\ref{centrality} shows the uncorrected charged
particle multiplicity in the TPC for these events and indicates the nine centrality bins
used in the analysis.  

To select charged particles with high efficiency and to minimize the
contribution from photon conversions and secondary charged particles produced
in the detector material, the following track requirements were applied
for tracks measured with the ITS and TPC. The tracks are required to
have at least 70 reconstructed space points out of the maximum 159 in the TPC and a
$\mean{\chi^2}$ per TPC cluster $\le 4$ 
(with two degrees of freedom per cluster). 
Additionally, at least two of the
six ITS layers must have a hit associated with the track.
Tracks are rejected if their distance of closest approach to the primary vertex is
larger than 0.3 cm in the transverse plane and 0.3 cm in
the longitudinal direction.  For the selected
tracks the reconstruction efficiency and remaining contamination is
estimated by Monte Carlo simulations of HIJING~\cite{refHIJING}
and Therminator~\cite{Kisiel:2005hn,Bozek:2010wt} events
using a GEANT3~\cite{geant3} detector simulation and event reconstruction.
The reconstruction efficiency for tracks with $0.2 < p_{\rm t} <
1.0$ GeV/$c$ increases from 60 to 70\% after which it stays constant at 
$70 \pm 5$\%. The contamination from secondary interactions and photon conversions 
is less than 5\% for $p_{\rm t} = 0.2$ GeV/$c$ 
and less than 1\% for $p_{\rm t} > 1$ GeV/$c$. 
Both the efficiency and 
contamination as a function of transverse momentum do not change significantly
as a function of multiplicity and are therefore the same for all
centrality classes.  

An alternative analysis was performed with tracks 
reconstructed using only the TPC information. 
For these tracks the same selections were applied except for the requirement of hits in the ITS and 
allowing for a  larger closest distance to the primary vertex, smaller than 3.0 cm in the transverse plane and 3.0 cm in
the longitudinal direction.
The reconstruction efficiency for these 
tracks with $0.2 < p_{\rm t} < 1.0$ GeV/$c$ increases from 
70 to 80\% after which it stays constant at $80 \pm 5$\%. 
The contamination is less than 6\% at $p_{\rm t} = 0.2$ GeV/$c$ 
and drops below 1\% at $p_{\rm t} > 1$ GeV/$c$. 
For this track selection the efficiency and contamination as a function of
transverse momentum also do not depend significantly on the track density
and are therefore the same for all centrality classes. The relative
momentum resolution for tracks used in this analysis was better than
5\%, both for the combined ITS--TPC and TPC-standalone tracks.
The results obtained from the ITS-TPC and TPC standalone tracking are in 
excellent agreement. Due to the smaller corrections for the azimuthal 
acceptance, the results obtained using the TPC standalone tracks are 
presented in this Letter.
 
\begin{figure}[thb]
  \begin{center}
    \includegraphics[width=0.49\textwidth]{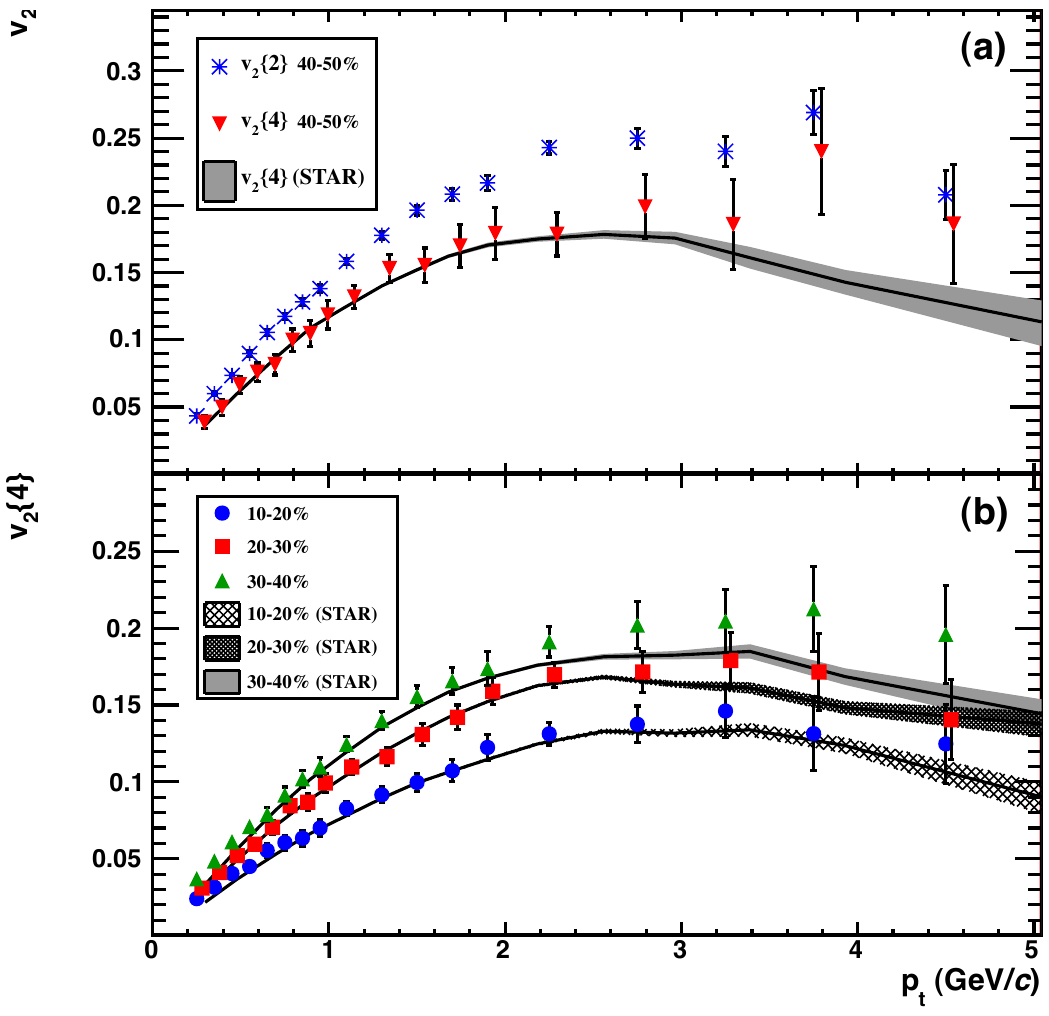}
    \caption{
    (color online)
    a) $v_2$(\pt) for the centrality bin 40--50\% from the 2- and 4-particle
    cumulant methods for this measurement and for 
    Au--Au collisions at $\snn~=~200$~GeV.
    b)  $v_2$\{4\}(\pt) for various centralities compared to STAR
    measurements. The data points in the 20--30\% centrality bin 
    are shifted in $p_{\rm t}$ for visibility.}
    \label{diff_flow}
  \end{center}
\end{figure} 
The $p_{\rm t}$-differential flow was measured
for different event centralities using various analysis techniques.
In this Letter we report results obtained with 2- and 4-particle
cumulant methods~\cite{Borghini:2001vi}, denoted $v_2\{2\}$ and
$v_2\{4\}$. 
To calculate multiparticle cumulants we used a new  
 fast and exact implementation~\cite{Bilandzic:2010jr}. 
The $v_2\{2\}$ and $v_2\{4\}$ measurements
have different sensitivity to flow fluctuations and {\it nonflow effects} 
-- which are uncorrelated to the initial geometry. Analytical
estimates and results of simulations show that nonflow contributions
to $v_2\{4\}$ are negligible~\cite{Adler:2002pu}.  
The contribution from flow fluctuations
is positive for $v_2\{2\}$ and negative for $v_2\{4\}$~\cite{Miller:2003kd}.  
For the integrated elliptic flow we also fit the 
flow vector distribution~\cite{Adams:2004wz} and use the Lee-Yang 
Zeroes method~\cite{Bhalerao:2003xf}, 
which we denote by $v_2\{$q-dist$\}$ and 
$v_2\{{\rm LYZ}\}$, respectively~\cite{Voloshin:2008dg}.  
In addition to comparing the 2- and 4-particle cumulant results we also
estimate the nonflow contribution by comparing to correlations of 
particles of the same charge. 
Charge correlations due to processes
contributing to nonflow (weak decays, correlations due to jets,
etc.) lead to stronger correlations between particles of unlike 
charge sign than like charge sign.

Figure~\ref{diff_flow}a shows $v_2$(\pt) for the centrality class 
40--50\% obtained with different methods. For comparison, we
present STAR measurements~\cite{yuting,:2008ed} for the same
centrality from Au--Au collisions at
$\snn~=~200$~GeV, indicated by the shaded area.  
We find that the value of $v_2$(\pt) does not change within uncertainties 
from $\snn~=~200$~GeV to 2.76~TeV.
Figure~\ref{diff_flow}b presents $v_2$(\pt) obtained with
the 4-particle cumulant method for three different centralities,
compared to STAR measurements.
The transverse momentum dependence is qualitatively similar for all
three centrality classes. 
At low $p_{\rm t}$ there is agreement of  $v_2$(\pt) with STAR data within uncertainties.

\begin{figure}[thb]
 \begin{center}
   \includegraphics[width=0.49\textwidth]   {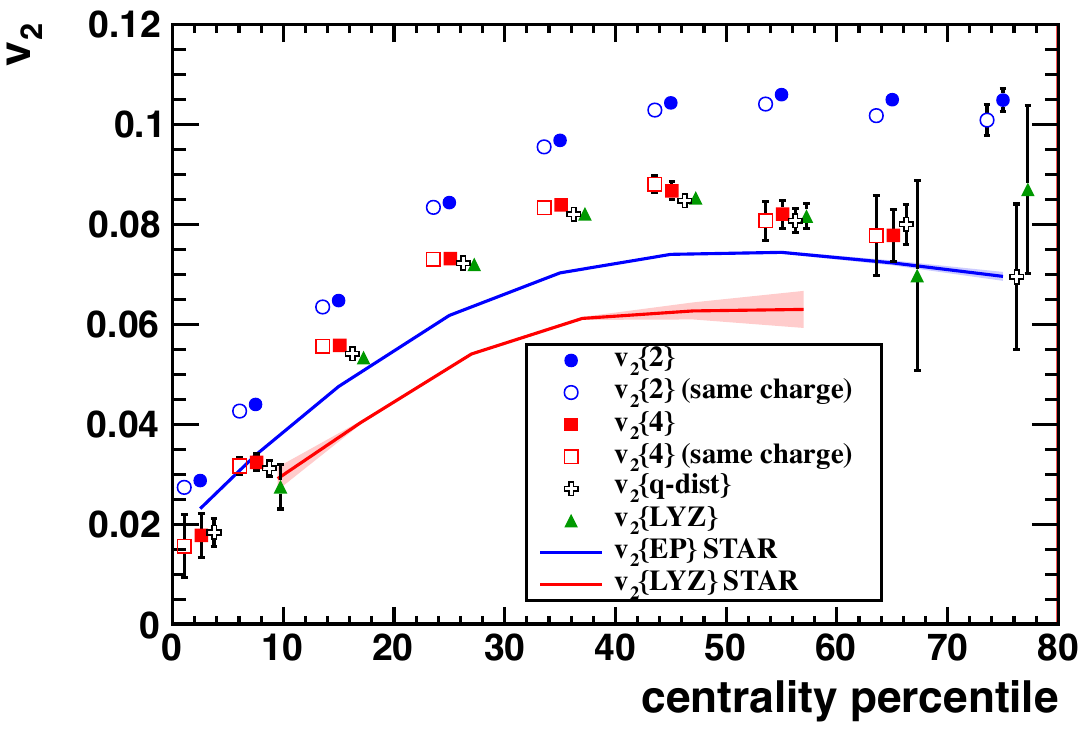}
   \caption{ 
   (color online)
   Elliptic flow integrated over the $p_{\rm t}$ range $0.2<p_{\rm t}<5.0$~GeV/$c$, as a function of event centrality, 
   for the 2- and 4-particle cumulant methods, a fit of 
   the distribution of the flow vector, and the Lee-Yang Zeroes method. 
   For the cumulants the measurements are shown for all charged particles (full markers) and 
   same charge particles (open markers). 
   Data points are shifted for visibility.
   RHIC measurements for Au--Au at $\snn~=~200$~GeV, 
   integrated over the $p_{\rm t}$ range $0.15<p_{\rm t}<2.0$~GeV/$c$, 
   for the event plane $v_2\{{\rm EP}\}$ and Lee-Yang Zeroes are shown by the solid curves. 
   }
    \label{int_flow}
 \end{center}
\end{figure}

The integrated elliptic flow is calculated for each centrality class using
the measured $v_2$(\pt) together with the charged particle $p_{\rm t}$-differential yield.  
For the determination of integrated elliptic flow the 
magnitude of the charged particle
reconstruction efficiency does not play a role. 
However, the relative
change in efficiency as a function of transverse momentum does
matter. 
We have estimated the correction to the integrated elliptic flow based on 
HIJING and Therminator simulations. 
Transverse momentum spectra in HIJING and Therminator 
are different, giving an estimate of the uncertainty in the correction.  
The correction is about 2\% with an uncertainty of 1\%.
In addition, the uncertainty due to  the centrality determination results in a relative uncertainty 
of about 3\% on the value of the elliptic flow.  

Figure~\ref{int_flow} shows that the integrated elliptic flow increases from central to 
peripheral collisions and reaches a maximum value in the 50--60\%
and 40--50\% centrality class of 0.106 $\pm$ 0.001 (stat) $\pm$ 0.004 (syst) 
and 0.087  $\pm$ 0.002 (stat) $\pm$ 0.003 (syst) for the 2- 
and 4-particle cumulant method, respectively.  
It is also seen that the measured integrated elliptic flow from the 4-particle cumulant, from fits of the 
flow vector distribution, and from the Lee-Yang Zeroes method, are in agreement. 
The open markers in Fig.~\ref{int_flow} show the results obtained for the cumulants
using particles of the same charge. The 4-particle cumulant results
agree within uncertainties 
for all charged particles and for the same charge particle data sets.
The 2-particle cumulant results, as expected due to nonflow, 
depend weakly on the charge combination. 
The difference is most pronounced for the
most peripheral and central events.  

\begin{figure}[htb]
 \begin{center}
   \includegraphics[width=0.49\textwidth]   {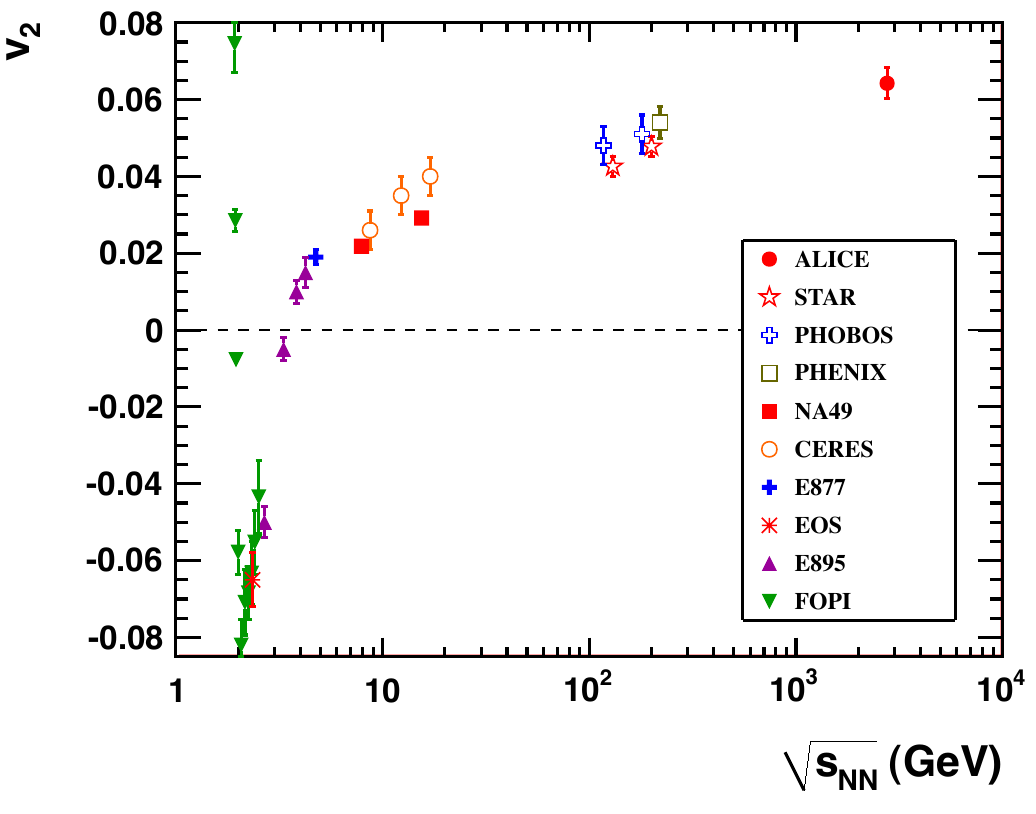}
   \caption{
   (color online) 
   Integrated elliptic flow at 2.76~TeV in \pb\ 20--30\% centrality class 
    compared with results from lower energies taken at
   similar centralities~\cite{Voloshin:2008dg, Andronic:2004cp}. 
   }
   \label{energy_dep}
 \end{center}
\end{figure}
The integrated elliptic flow measured in the 20--30\% centrality class
is compared to results from lower energies in
Fig.~\ref{energy_dep}. 
For the comparison we have corrected the integrated elliptic flow for the 
$p_{\rm t}$ cutoff of 0.2 GeV/$c$. The estimated magnitude of this correction is 
12 $\pm$ 5\% based on calculations with Therminator. 
The figure shows that there is a continuous increase in 
the magnitude of the elliptic flow for this centrality region
from RHIC to LHC energies.  
In comparison to the elliptic flow measurements in \au\ collisions at
$\snn~=~200$~GeV we observe about
a 30\% increase in the magnitude of $v_2$ at $\snn~=~2.76$~TeV. 
The increase of about 30\% is larger  than in current ideal 
hydrodynamic calculations at LHC multiplicities~\cite{Kestin:2008bh} but is in agreement with some 
models that include viscous corrections which at the LHC become 
less important~\cite{Masui:2009pw,Luzum:2009sb,Hirano:2010jg,Hirano:2005xf,Drescher:2007uh}.

In summary we have presented the first elliptic flow measurement at
the LHC.  The observed similarity at RHIC and the LHC of $p_{\rm t}$-differential elliptic flow 
at low $p_{\rm t}$ is consistent with
predictions of hydrodynamic models~\cite{Niemi:2008ta,Kestin:2008bh}.
We find that the integrated elliptic flow increases about 30\% from
$\snn~=~200$ GeV at RHIC to $\snn~=~2.76$~TeV. 
The larger integrated elliptic flow at the LHC is caused by the
increase in the mean $p_{\rm t}$. Future elliptic flow measurements 
of identified particles will clarify the role of 
radial expansion in the formation of elliptic flow.

\section{acknowledgements}
The ALICE collaboration would like to thank all its engineers and technicians for their invaluable contributions to the construction of the experiment and the CERN accelerator teams for the outstanding performance of the LHC complex.
The ALICE collaboration acknowledges the following funding agencies for their support in building and
running the ALICE detector:
Calouste Gulbenkian Foundation from Lisbon and Swiss Fonds Kidagan, Armenia;
Conselho Nacional de Desenvolvimento Cient\'{\i}fico e Tecnol\'{o}gico (CNPq), Financiadora de Estudos e Projetos (FINEP),
Funda\c{c}\~{a}o de Amparo \`{a} Pesquisa do Estado de S\~{a}o Paulo (FAPESP);
National Natural Science Foundation of China (NSFC), the Chinese Ministry of Education (CMOE)
and the Ministry of Science and Technology of China (MSTC);
Ministry of Education and Youth of the Czech Republic;
Danish Natural Science Research Council, the Carlsberg Foundation and the Danish National Research Foundation;
The European Research Council under the European Community's Seventh Framework Programme;
Helsinki Institute of Physics and the Academy of Finland;
French CNRS-IN2P3, the `Region Pays de Loire', `Region Alsace', `Region Auvergne' and CEA, France;
German BMBF and the Helmholtz Association;
Hungarian OTKA and National Office for Research and Technology (NKTH);
Department of Atomic Energy and Department of Science and Technology of the Government of India;
Istituto Nazionale di Fisica Nucleare (INFN) of Italy;
MEXT Grant-in-Aid for Specially Promoted Research, Ja\-pan;
Joint Institute for Nuclear Research, Dubna;
 %
National Research Foundation of Korea (NRF);
CONACYT, DGAPA, M\'{e}xico, ALFA-EC and the HELEN Program (High-Energy physics Latin-American--European Network);
Stichting voor Fundamenteel Onderzoek der Materie (FOM) and the Nederlandse Organisatie voor Wetenschappelijk Onderzoek (NWO), Netherlands;
Research Council of Norway (NFR);
Polish Ministry of Science and Higher Education;
National Authority for Scientific Research - NASR (Autoritatea Na\c{t}ional\u{a} pentru Cercetare \c{S}tiin\c{t}ific\u{a} - ANCS);
Federal Agency of Science of the Ministry of Education and Science of Russian Federation, International Science and
Technology Center, Russian Academy of Sciences, Russian Federal Agency of Atomic Energy, Russian Federal Agency for Science and Innovations and CERN-INTAS;
Ministry of Education of Slovakia;
CIEMAT, EELA, Ministerio de Educaci\'{o}n y Ciencia of Spain, Xunta de Galicia (Conseller\'{\i}a de Educaci\'{o}n),
CEA\-DEN, Cubaenerg\'{\i}a, Cuba, and IAEA (International Atomic Energy Agency);
The Ministry of Science and Technology and the National Research Foundation (NRF), South Africa;
Swedish Reseach Council (VR) and Knut $\&$ Alice Wallenberg Foundation (KAW);
Ukraine Ministry of Education and Science;
United Kingdom Science and Technology Facilities Council (STFC);
The United States Department of Energy, the United States National
Science Foundation, the State of Texas, and the State of Ohio.

\end{document}